\input{aipcheck}
\documentclass[
    ,final            
  ]
  {aipproc}

\layoutstyle{8x11single}

\usepackage{graphicx}
\usepackage{amsmath}
\usepackage{amssymb}
\newcommand{\dd}{\mathrm{d}} 
\newcommand{\Mpl}{M_\mathrm{pl}} 
\newcommand{\mpl}{m_\mathrm{pl}} 

\newcommand{\rr}{\mathrm}

\newcommand{\ueff}{\mathrm{eff}}
\newcommand{\uini}{\mathrm{ini}}
\newcommand{\ui}{\mathrm{i}}
\newcommand{\Veff}{V_\ueff}

\newcommand\doingARLO[2][]{%
  \ifx\mmref\undefined #1\else #2\fi
}

\begin{document}

\title{Anamorphosis in hybrid inflation:  \\ How to avoid fine-tuning of initial conditions?}


\author{S\'ebastien Clesse} {address={Service de Physique Th\'eorique, Universit\'e Libre de
  Bruxelles, CP225, Boulevard du Triomphe, \\ 1050 Brussels, Belgium},
altaddress={Center for Particle Physics and Phenomenology, Louvain
  University, 2 chemin du cyclotron, \\ 1348 Louvain-la-Neuve, Belgium},email=seclesse@ulb.ac.be }

\begin{abstract}
In order to generate more than 60 e-folds of accelerated expansion in original hybrid inflation, 2-fields trajectories are usually required to be initially fine-tuned in a very narrow band along the inflationary valley or in some isolated points outside it.   From a more precise investigation of the dynamics, these points which can cover a non-negligible proportion of the space of sub-planckian initial field values, depending on the potential parameters, are shown to be organised in connected domains with fractal boundaries.  They correspond to trajectories first falling towards the bottom of the potential, then climbing and slow-rolling back along the inflationary valley.   The full parameter space, including initial velocities and all the potential parameters, is then explored by using Monte-Carlo-Markov-Chains (MCMC) methods.  Results indicate that successful initial conditions (IC) outside the valley are not localized in the parameter space and are the dominant way to realise inflation, independently of initial field velocities.   Natural bounds on parameters are deduced.  The genericity of our results is confirmed in 5 other hybrid models from various framework.         
\end{abstract}

\maketitle

\section{Introduction}

\ \ \ \ \  Inside the jungle of  inflationary models, the hybrid class is particularly promising because easily embedded in high energy frameworks like supersymmetry (SUSY), supergravity (SUGRA) or grand unification theories (GUT).  

In original hybrid model~\cite{Linde:1993cn,Copeland:1994vg}, the inflaton is coupled to an auxiliary field, and inflation is assumed to occur in a nearly flat valley of the potential, before ending due to an Higgs-type instability.   If the viability of the model can be discussed\footnote{A disfavoured blue spectrum of initial perturbations is predicted in usual effective 1-field slow-roll approximation of the model.  However, as discussed in~\cite{Clesse:2008pf}, the spectrum can be red in particular cases, or when formation of cosmic strings at the end of inflation is taken acount~\cite{Bevis:2007gh}.} when confronted to Cosmic Microwave Background (CMB) observations, it is nevertheless a very interesting toy model, with a dynamic similar to many other hybrid realisations of inflation in various high energy frameworks.

For field values smaller than the reduced Planck mass\footnote{Natural limit if the model results from SUSY~\cite{Mendes:2000sq}}, the original hybrid model was found to suffer from a fine-tuning problem of initial field values. It is related to the necessity for the trajectories to start either in a very narrow band along the inflationary valley, either in some apparently isolated points outside this valley, in order to solve the standart cosmological problems (e.g. horizon problem).   

However, some inconcistencies can be pointed out by comparing grids of initial conditions plotted by Tetradis~\cite{Tetradis:1997kp}, and afterwards by Mendes and Liddle~\cite{Mendes:2000sq} with a higher resolution, for similar sets of parameters.   Some questions remained unanswered:  are successful points outside the inflationary valley isolated or do they form some structure?  Which proportion of the initial field space do they occupy?   What is the origin of these points and what are the effects of the potential parameters on these successful initial conditions?

Those arguments have led to the re-analysis of the problem~\cite{Clesse:2008pf}.  The successful trajectories starting outside the inflationary valley have been found to present a specific behaviour:  after a phase of fast roll towards the bottom of the potential and some oscillations around it, they become oriented along the valley and climb it before slow rolling back along it and generating a large number of e-folds.  These initial conditions occupy a non negligible part of the initial field space (up to 20 \%) for many sets of parameters, potentially resolving the above mentioned fine-tuning problem.    Moreover, they are organised in a complex structure whose fractal properties can be studied~\cite{Clesse:2009ur}.  

In order to verify the robustness of these observations in the whole parameter and initial condition (IC) space, including the possibility of non-vanishing initial velocities, a Monte-Carlo-Markov-Chains (MCMC) analysis has been performed~\cite{Clesse:2009ur}.   A remarkable result of this method is the establishment of new natural bounds on potential parameters based only on the requirement of a sufficiently long inflationary era and the absence of fine-tuning of initial fields values.

Finally, the problem of initial conditons in other hybrid models from various framework (supersymmetry, supergravity, extra-dimentions) have been studied~\cite{Clesse:2008pf}, with the conclusion that successful initial field values outside the valley(s), called \textit{anamorphosis points} by analogy to the optical phenomena, is the generic and more probable way to generate inflation.

\section{Original hybrid model and its dynamics}

The original hybrid model was proposed in
Refs.~\cite{Linde:1993cn,Copeland:1994vg}, its potential reads
\begin{equation} \label{eq:potenhyb2d}
V(\phi,\psi) = \frac{1}{2} m^2 \phi^2 + \frac {\lambda}{4} \left(\psi^2
- M^2 \right)^2 +\frac{\lambda'}{2} \phi^2 \psi^2.
\end{equation}
The field $\phi$ is the inflaton and $\psi$ is the auxiliary
Higgs-type field while $\lambda$, $\lambda'$ are two positive coupling
constants and $m$, $M$ are two mass parameters.  
This potential presents a nearly flat valley, along $\psi = 0$, in which inflation is assumed to be realized in the
false-vacuum, before ending due to a tachyonic instability when the inflaton reaches a critical value $\phi_{\rr c} = M
\sqrt{\frac {\lambda}{\lambda'}}$.  From this point, the system would classically evolve toward
one of its true minima $\phi=0$, $\psi =\pm
M$.

In order to observe the effects of the parameters values on the dynamics, it is convenient to rewrite the potential into the form
\begin{equation} \label{eq:potenhyb2dNEW}
V(\phi,\psi) = \Lambda^4 \left[  \left( 1 - \frac{\psi^2}{M^2} \right)^2 
+ \frac{\phi^2}{\mu^2} + \frac{\phi^2 \psi^2}{\nu^4}\right] ,
\end{equation}
in which $M,\mu,\nu$ are three mass parameters. With this
expression, the critical point of instability now reads
\begin{equation}
\label{eq:critical}
\phi_{\rr c} = \sqrt 2  \frac{\nu^2}{M}\,.
\end{equation}
When trajectories are slow-rolling along the valley, the potential can be restricted to the usual effective one-field form
\begin{equation}
  \Veff(\phi) = \Lambda^4 \left[ 1 + \left( \frac \phi \mu \right)^2  \right].
\end{equation}   

In a flat Friedmann--Lema\^{\i}tre-Robertson--Walker metric, the
equations governing the two-fields dynamics are the
Friedman-Lema\^{\i}tre equations
\begin{equation} \label{eq:FLtc12field}
\begin{split}
H^2 &= \frac {8\pi }{3 \mpl^2}  \left[ \frac 1 2 \left(\dot
\phi^2 + \dot \psi^2 \right)  + V(\phi,\psi) \right], \\
\frac{\ddot a }{a} &= \frac {8\pi}{3 \mpl^2} \left[ - \dot \phi^2
- \dot \psi^2 + V(\phi,\psi ) \right],
\end{split}
\end{equation}
as well as the Klein-Gordon equations 
\begin{equation} \label{eq:KGtc2field}
\begin{split}
&\ddot \phi + 3 H \dot \phi + \frac {\partial
V(\phi,\psi)}{\partial \phi} = 0, \\
&\ddot \psi + 3 H \dot \psi + \frac {\partial
V(\phi,\psi)}{\partial \psi} = 0,
\end{split}
\end{equation}
where $\mpl \simeq 1.2 \times 10^{19}\; \mathrm{GeV} $ is the Planck
mass\footnote{Throughout the paper, $\mpl$ denotes the physical Planck
  mass, and $\Mpl$ stands for the reduced Planck mass $\Mpl \simeq 0.2
  \mpl\simeq 2.4\times 10^{18}$~GeV.}, $H \equiv \dot a/a$ is the
Hubble parameter, $a$ is the scale factor and a dot means derivative
with respect to cosmic time.

In order to test trajectories from the space of initial conditions, these equations have been integrated numerically.  In this paper, we assume that successful inflation means a generation of at least  $N=\ln(a/a_\uini)\simeq 60$ e-folds before the time when the trajectory is trapped definitively by one of the two global minima, corresponding to the condition
\begin{equation}
\frac 1 2 \left( \dot \phi^2 + \dot \psi^2 \right) + V(\phi,\psi) < V(0,0) = \Lambda^4 .
\end{equation}

\section{Space of successful initial conditions}

\subsection{Anamorphosis points}

\begin{figure}[]  
  \includegraphics[width=9cm]{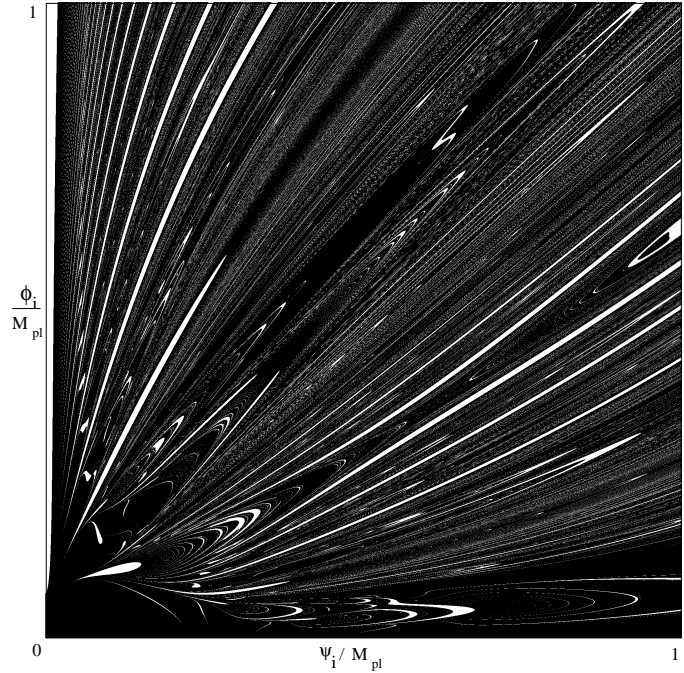} 

  \caption{Grid (2040x2040 points) of initial field values producing more/less than $60$ e-folds of
    inflation (white/black regions). The potential parameters are set
    to $M=0.03 \  \mpl$, $\mu = 636 \ \mpl$, $\nu^2 = 3 \times
    10^{-4}\mpl^2$. }
  \label{fig:grid}
\end{figure}

High resolution grids of initial field values at fixed initial velocities and potential parameters~\cite{Clesse:2008pf} indicate that successfull points outside the valley are oganised in a complex structure of thin lines and crescants (Fig. \ref{fig:grid}), instead of being isolated as it was previously found in~\cite{Mendes:2000sq}.  Moreover, these regions can occupy up to 20\% of the space of initial conditions, depending on the chosen set of potential parameters.  

\begin{figure}[]  
   \includegraphics[width=9cm]{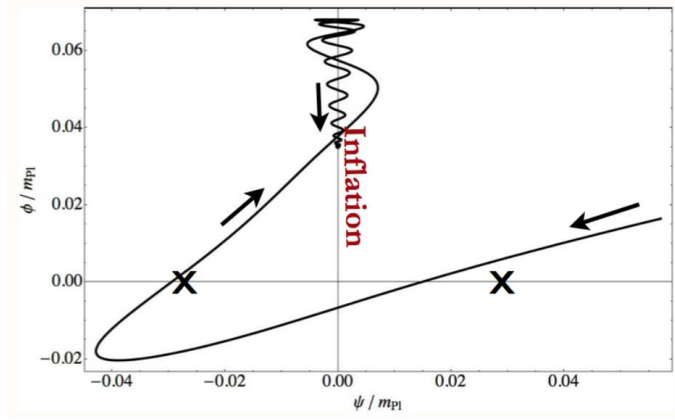}

  \caption{Typical behaviour of successful trajectory starting outside the inflationary valley, first falling towards the bottom of the potential, then oscillating and climbing the inflationary valley ($\psi =0$ direction) instead of being trapped by one of the global minima of the potential (represented by crosses).  Inflation occurs when it slow-rolls back along the valley.}
  \label{fig:trajectory}
\end{figure}

Such trajectories have a characteristic behaviour.  They first fall in the bottom of the potential, but after some rebonds and oscillations on its sides, instead of being trapped by one of the global minima, they become oriented in the direction of the inflationary valley, climb it, and then slowly roll back along it and generate inflation (Fig. \ref{fig:trajectory}).  Thereby it is possible to identify critical points along these trajectories corresponding to the instant when the trajectory stops to climb the valley and starts to slow-roll back.  On each of them, the velocity of the trajectory is (quasi-)vanishing and because they lead to successful inflation, they are located inside the successful band in the valley.   Therefore, there exists a mapping between each successful initial point  and  points with (quasi-)vanishing velocity inside the valley.  The intricated structure formed by successful points can thus be seen as a distorded image of the inflationary valley.  We called it \textit{anamorphosis} with the potential playing a role similar to a curved mirror, by analogy to the optical phenomena in which an anamorphosis is a deformed image of a concrete object seen through a deforming mirror.  

\subsection{Fractal properties}

Looking attentively at Fig.\ref{fig:grid}, the question of the self-reprocuctivity of the structures at different scales can be tackled.  This question is related to the fractal properties of the set of successful initial field values.  It is an important issue because a fractal behaviour can induce problems in the correct mathematical definition of a measure on the space, and by extension on the way to calculate the relative area covered by successfull points.   

The dynamical system owns three attractors:  two of them are the global minima of the potential, the third in the inflationary valley.  Successful initial field values outside the valley form the basin of attration of the valley.  Since this attractor is a dense set of dimension 2 and since the mapping between successful initial field values and the valley is continuous, one expects the successful initial field values outside the valley to form a dense set of dimension 2~\cite{Falconer:fracgeo}.

A numerical calculation of the box-counting \footnote{B.C. dimension can be set equal to the Hausdorff fractal dimension, as discussed in~\cite{Clesse:2009ur}} (B.C.) fractal dimension of the strucure boundaries and of its surface itself have been performed~\cite{Clesse:2009ur}.   Results verify that anamorphosis structure have a non-fractal convergent area (B.C. dimension $D_a = 2.0 $)  but indicate that boundaries are fractal (e.g. B.C. dimension $D_b= 1.20$ for set of parameters $M=0.03 \  \mpl, \lambda=\lambda'=1, m=10^{-6} \mpl$, vanishing initial velocities, and initial field values in the range $0 < \phi, \psi < \Mpl$).  These properties similar to the well-known Mandelbrot set, allow to use the usual Lebesgues measure on the space and to simply count the proportion of successful points in grids of initial conditions to determine the relative surface of the anamorphosis structure.

\subsection{MCMC statistical analysis}

Until now are discussed the properties of the space of initial field values for some slices in the potential parameter space and only for vanishing initial velocities.  If the possibility to avoid fine-tuning of initial conditions through anamorphosis points have been introduced for particular sets of parameters, it is fundamental to study it in the whole parameter space in order to determine if this observation stay valid or if it is a local phenomena appearing for only a very restricted range of parameters.  

In order to clarify these remarks, a statistical exploration of the whole potential parameter and initial conditions space (7-th dimensional space) has been performed.  We have defined a probability measure on that space, and using Bayesian inference, one can asses the posterior probability distributions of all parameters to get more than 60 e-folds of accelerated expansion.  A Monte-Carlo-Markov-Chains method have been used to estimate these distributions.   Technical details about the algorithm used to construct the Markov-Chains and the way to derive posterior probability distributions  can be found in~\cite{Clesse:2009ur}.

\subsubsection{Priors}
In order to explore identically all scales of parameters values, we have chosen a flat prior for the logarithm of the parameters.   Like small initial velocities are expected to not change dramatically the dynamics, a flat prior on initial $\frac {\dd \phi}{\dd N}$ and $\frac{\dd \psi}{\dd N}$ have been chosen, within the natural constraint
\begin{equation}
  v^2 \equiv  \left(  \frac{\dd \phi} {\dd N} \right)^2 + \left( \frac{\dd
      \psi} {\dd N} \right)^2 < \frac{6}{\Mpl^2}.
\end{equation}

This upper bound is derived easily from the equations of motion (\ref{eq:FLtc12field},\ref{eq:KGtc2field}).  Finally, we have also chosen a flat prior on initial field values. 


\subsubsection{Posterior probability distributions}

\begin{figure*}[]
\includegraphics[width=14cm]{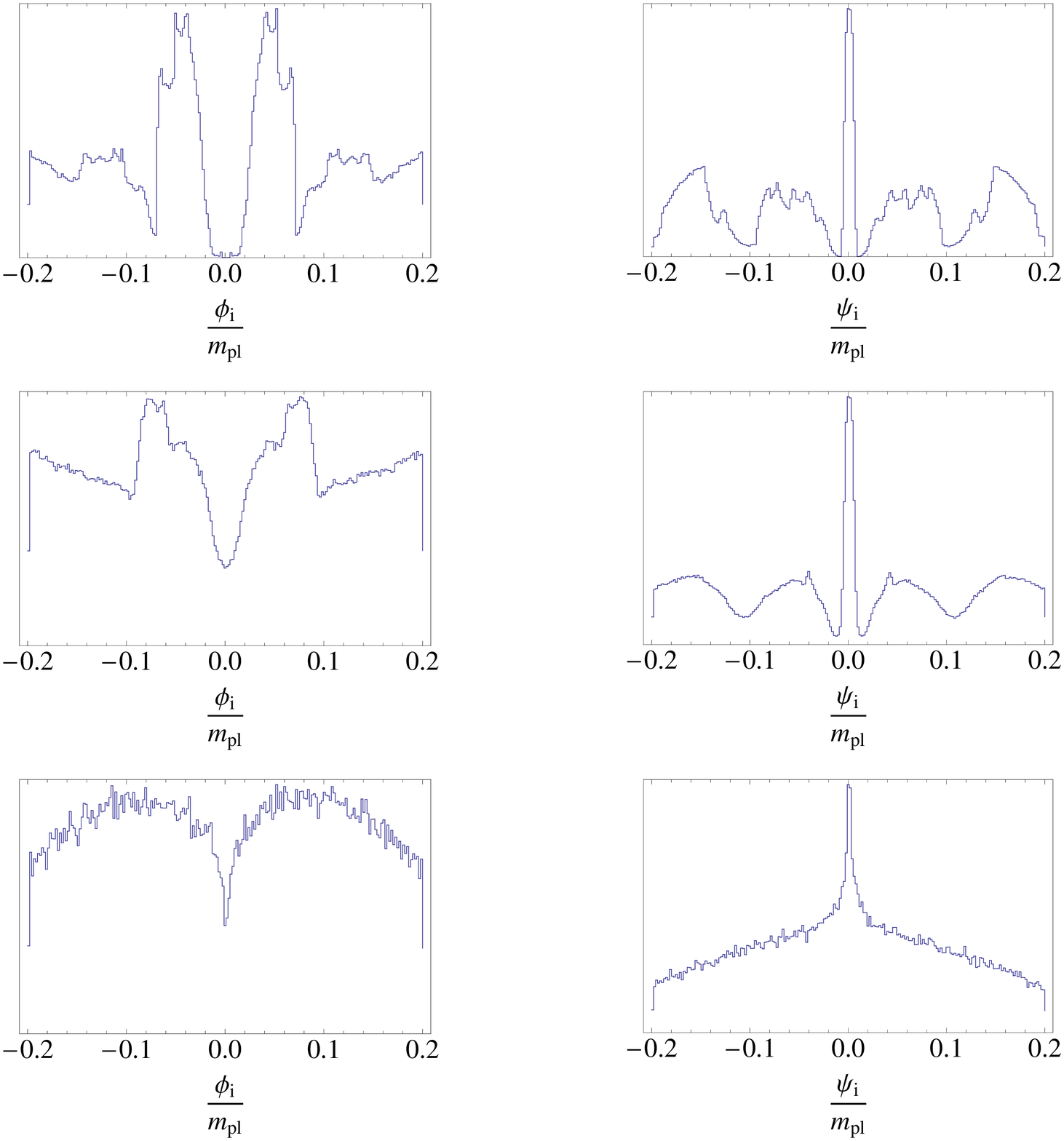}
\caption{Marginalised posterior probability distributions for the
  initial field values $\phi_\ui$ and $\psi_\ui$. The top panels
  correspond to vanishing initial velocities and fixed potential
  parameters, the middle ones are marginalised over velocities at
  fixed potential parameters, while the lower panels are marginalised
  over velocities and all the potential parameters.}
\label{fig:1Dfields}
\end{figure*}

Posterior probability distributions of the fields (Fig.\ref{fig:1Dfields})   indicate that anamorphosis trajectories are the most probable way to generate inflation, compared to trajectories starting in the valley.  This result stands when distributions are marginalised over the initial velocity space, as well as in the whole parameter space.  

\begin{figure}[]
\includegraphics[width=60mm]{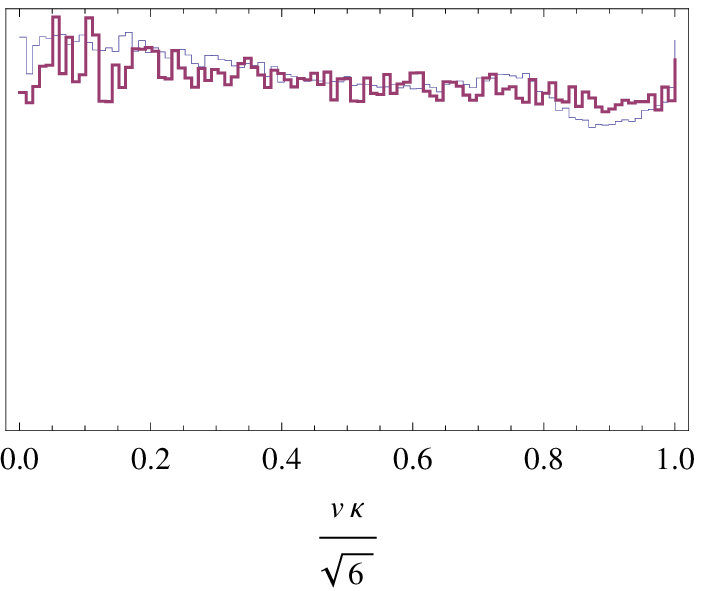}  
\raisebox{3.5mm}{
\includegraphics[width=60mm]{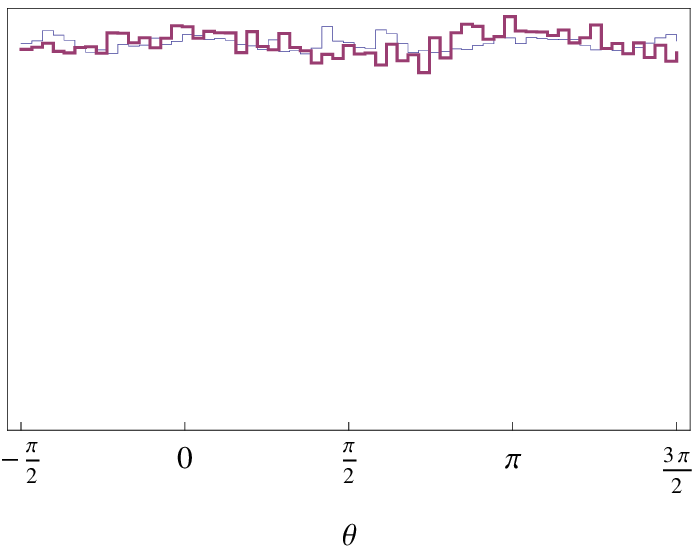} }
\caption{Marginalised posterior probability distributions for the
  modulus (top) and angle (bottom) of initial field velocity. The thin
  curves are obtained at fixed potential parameters, while the
  thick are obtained after a full marginalisation over all the model
  parameters. As expected from Hubble damping, all values are
  equiprobable since the field do not keep memory of the initial
  velocity.}
\label{fig:1Dspeeds}
\end{figure}

Moreover,  posterior probability distributions for the intensity and orientation of the initial velocity vector are flat (see Fig.\ref{fig:1Dspeeds}).   We can therefore conclude that adding the possibility of initial velocities \emph{do not} change the probability to generate a sufficient period of inflation, because of the Hubble damping in the Klein-Gordon equations (\ref{eq:KGtc2field}).   Actually, due to this friction term, only a small fraction of e-folds can be generated before the velocities reach the slow-roll attractor, whatever their initial values.

\begin{figure}
\includegraphics[width=6cm]{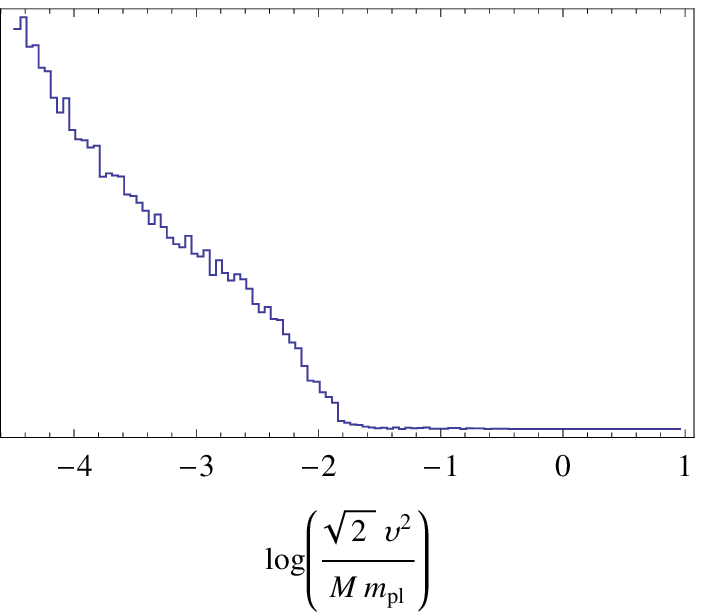}
\raisebox{5.5mm}{
\includegraphics[width=6cm]{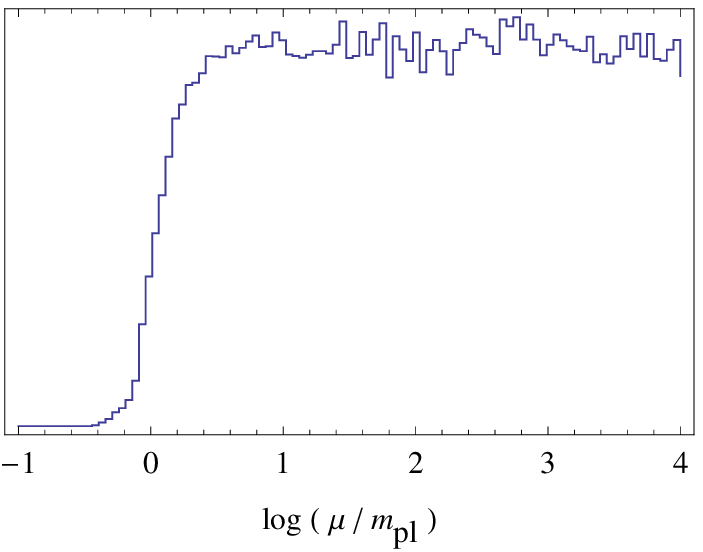}}
\caption{Marginalised posterior probability
  distribution for the combinaison of parameters $\nu^2/(M \Mpl)$ (left), and for paramter $\mu$ (right).  A respectively upper/lower  bound can be deducted on these parameters. }
\label{fig:1Dinstpt}
\end{figure}

Secondly, the probability distributions of the potential parameters show that the successful initial field values outside the valley are generic in a large part of the whole parameter space.   We obtain a natural upper bound on the combination of parameters 
\begin{equation} 
\sqrt 2 \frac{\nu^2}{M} < 0.004 \ \mpl \ 95\%\  \mathrm{C.L.}
\end{equation}
 corresponding to the position of the instability point inside the inflationary valley (see Fig\ref{fig:1Dinstpt} left).   The value of the inflaton at the instability point is thus smaller than the reduced Planck mass, and the last 60 e-folds of inflation are sure to be generated in a sub-planckian regime inside the valley
 
Moreover, a lower bound on the parameter $\mu$ have been obtained (Fig\ref{fig:1Dinstpt} right).   
\begin{equation}
\mu > 1.7 \ \mpl \ 95\% \ \mathrm{C.L.}
\end{equation}
This bound is non-trivial because it corresponds to the apparition of slow-roll violations in the dynamics of the trajectories inside the inflationary valley.   These violations can lead to the disappearance of the small field phase expected if the slow-roll formalism is used, as discussed in~\cite{Clesse:2008pf}.

Therefore, from MCMC analysis and the resulting posterior probability density distributions of initial conditions and parameters, one can not only determine that in a large part of the parameter space, successful trajectories starting outside the valley are natural, whatever the initial velocities, but also to show that natural bounds on parameters can be established, based only on the requierement of 60 e-folds of inflation.   This is thus a new promising way to constrain naturally hybrid models of inlfation, and their related physics. 

\section{Other hybrid models}

Original hybrid model is a good toy model for other hybrid realisations of inflation in various framework, which present a similar dynamics.   It is therefore interesting to test the robustness of our results, in some more realistic\footnote{realistic in the double sense that these models are in accord with CMB datas and are embedded in a high energy framework} hybrid models.   

We have first calculated grids of initial field values for many sets of parameters, for \textit{supersymmetric smooth}~\cite{Lazarides:1995vr,Lazarides:2007fh} \textit{hybrid model, supersymmetric shifted hybrid model}~\cite{Jeannerot:2000sv,Jeannerot:2002wt}, for their supergravity versions, as well as for \textit{Radion Assisted Gauge inflation}~\cite{Fairbairn:2003yx}.   For all these models, a non negligible part of the initial field space have been found to be successful outside the central valley(s), with up to 80 \% of successful points for \textit{smooth hybrid inflation}~\cite{Clesse:2008pf}.    

Therefore, the observation that fine-tuning problem can be avoided due to the presence of anamorphosis points stays valid for those five hybrid models.   

\begin{figure}[h!]
\includegraphics[width=10cm]{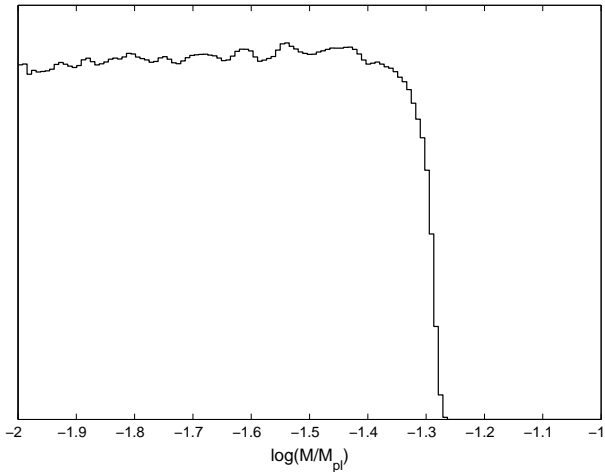}
\caption{Marginalised posterior probability distribution of the mass scale $M$ in F-term SUGRA inflation}
\label{fig:fsugra}
\end{figure}

A similar MCMC method have been applied to F-term hybrid model in supergravity~\cite{Clesse:2009ur}.  In that particular model, the potential contains only one relevant parameter $M$.   Similar results have been obtained for probability density distributions of initial field values and velocities.   Moreover, from probability density distribution of $M$ (Fig.\ref{fig:fsugra}), a natural upper bound 
\begin{equation}
\log (M/\Mpl)  < -1.33  \ 95\% \ \mathrm{C.L.}.
\end{equation}
has been established on the unique parameter of the model.

\section{Conclusion}

For original hybrid model, we have shown that there is no need of a mechanism to fine-tune the sub-planckian initial conditions of the fields along the inflationary valley in order to generate a sufficiently long inflationary era.  Actually, depending on potential parameters, a non-negligible part of the space of initial field values (up to 20\%) outside the valley leads to sufficient inflation.  The statistical MCMC analysis of the initial field values, initial field velocities and parameter space has shown that this observation is true in a large but bounded part of the whole parameter space and do not depend on the initial velocity vector.   From their posterior probability density distributions, it is thus possible to derive new natural  bounds on parameters of hybrid models with the only requirement of a sufficiently long inflation. 

The behaviour of the successful trajectories starting outside the inflationary valley have been studied.  They are characterised by a phase of fast-roll towards the bottom of the potential, followed by a climbing phase of the inflationary valley.  Inflation occurs when the trajectory slow-rolls back along it.    These successful IC form the basin of attraction of the inflationary valley and exhibit a fractal boundary, in a way similar to the well-known Mandelbrot set.   

Moreover, this mechanism appears to be generic in five other hybrid models coming from various frameworks.  

Finally, the next step will be to extend our MCMC method in order to include constraints on parameters from CMB data on the power spectrum of adiabatic and isocurve perturbations, as well as from cosmic strings formation at the end of inflation, should be a very challenging but interesting perspective, in order to analyse completely the viability of hybrid models in the light of both theory and experimental data.

\begin{theacknowledgments}
This work has been done in collaboration with C. Ringeval and J. Rocher.  It is a pleasure to thank T. Carletti, A. Fuzfa, A. Lemaitre and M. Tytgat for fruitful discussions and comments. S.C. is supported by the Belgian Fund for research (F.R.I.A.). 
\end{theacknowledgments}

\doingARLO[\bibliographystyle{aipproc}]
          {\ifthenelse{\equal{\AIPcitestyleselect}{num}}
             {\bibliographystyle{arlonum}}
             {\bibliographystyle{arlobib}}
          }

\bibliography{biblio}

\end{document}